\documentclass[11pt,twoside]{article}

\usepackage{asp2006}
\usepackage{lscape}
\usepackage{graphicx}

\begin{document}
\title{Quest for truly isolated galaxies}   
\author{Noah Brosch}   
\affil{Wise Observatory and the Beverly and Raymond Sackler School of Physics and Astronomy, Tel Aviv University, Tel Aviv 69978, Israel}    

\begin{abstract} 
I describe attempts to identify and understand the most isolated galaxies starting from my 1983 Leiden PhD thesis, continuing through a string of graduate theses on various aspects of this topic, and concluding with an up-to-date account of the difficulty to find really isolated objects. The implication of some of the findings revealed on the way and presented here is that the nearby Universe may contain many small dark-matter haloes, and that some such haloes may possibly be accreting intergalactic gas to form dwarf galaxies.
\end{abstract}


\section{Introduction}   
Although Karachentseva (1973) was the first to define a sample of isolated galaxies (IGs), her work did not percolate into the community until the 1980s. Other attempts to define a sample of IGs were better known: e.g., a 12-object IG sample of Huchra \& Thuan (1977) whose galaxies have no companions with m$\leq$15.7 within 45 arcmin. This modest sample was studied multi-spectrally by Brosch (1983). In particular, multi-aperture UBV photoelectric photometry (Brosch \& Shaviv  1982) was combined with near-IR photometry (Brosch \& Isaacman 1982), 5 GHz radio continuum synthesis (Brosch \& Krumm 1984), and IUE UV spectroscopy (Brosch et al. 1984) to show that the Huchra \& Thuan IGs have enhanced star formation (SF) in their central regions.

In preparation for the Isolated Galaxies meeting in Granada I revisited these putative IG, searching for neighbouring galaxies within 3 Mpc and 300 km sec$^{-1}$ and using the method developed by Spector (these proceedings). I found that the objects were not isolated, some having more that 100 galaxies within the search region.

\section{What are IGs and why are they interesting?}

This finding, that galaxies presumed to be isolated are not so, begs the questions of what exactly does the term ``isolated'' mean for a galaxy, and why should IGs be studied. The isolation should allow the identification of galaxies least affected by interactions. Not only SF and the activation of an AGN might be affected by interactions, but also the morphology of a galaxy. It is possible to estimate the number of past interactions a galaxy might have had based on its size $r$, the velocity dispersion of  neighbourhood galaxies $\Delta v$, and the neighbourhood galaxy density $n$, by using $N \simeq (\frac{r}{100 \, {\rm kpc}})^2 \frac{v}{100 \, {\rm km \, sec^{-1}}}\frac{n}{(1 \, Mpc)^3}\,\frac{1}{H_0}$ to represent the number of possible past encounters during one Hubble time. A choice of typical parameters for sparse regions implies a negligible possibility of a past interaction.  Di Mateo et al. (2008) showed that most galaxy interactions or mergers trigger only moderate SF enhancements, but when studying IGs one would like to eliminate such cases as well.

The AMIGA IG selection, originally based on the Karachentseva (1973) catalog, was refined by purging galaxies that a variety of criteria showed not to be isolated. Sulentic et al. (2006) eliminated AMIGA galaxies with a damaged appearance based a visual check, since such objects may have had past interactions with unseen companions. However, Spector (unpublished) found that some AMIGA galaxies had HI companions. Some on the objects in another IGs sample (Pisano et al. 2002) show warps and optically-faint HI companions; they are definitely not isolated. ``Pure'' IGs are therefore rare, if they exist at all.

\section{Isolated galaxies are not alone}
Brosch et al. (2006) searched for star-forming companions in the vicinity of star-forming dwarf galaxies (DGs) by H$\alpha$ imaging. This method allows the detection of faint star-forming objects in the physical vicinity of a galaxy without requiring very deep spectroscopy. We selected five dwarf (M$\geq$--18 mag) galaxies at least 2.5 Mpc away from any other NED galaxy thus located in very under-dense regions, which served as search field centers (FCs). The DGs were selected to exhibit H$\alpha$ emission to indicate recent or on-going star formation, with the goal of understanding why are they presently forming stars, when large-scale SF triggers such as density waves are absent.

We used the WO H$\alpha$ filters, which have FWHM$\sim$50\AA\, transmission profiles, to separate adjacent $\sim1000$ km sec$^{-1}$ velocity slices in an imaging survey of the immediate region around each galaxy. One filter was selected at the approximate redshift of each FC galaxy and the others sampled higher or lower redshifts. We identified 20 possible faint SF neighbor galaxies, 17 not previously catalogued, in
three of the five search fields where we had good quality data. The objects were small, almost star-like, and were at a few 10s to a few 100s kpc projected distance from their FC galaxies but at approximately the same redshift, since they showed up in the same net-H$\alpha$ image as the FC object.

These small HII galaxies are similar to the H$\alpha$ knot found by Brosch et al. (2008) at a 63 kpc projected distance from AM1934-563, and to the similar objects identified by Kellar et al. (2008). The relatively large number of possible neighbour star-forming galaxies, combined with their relative faintness and small sizes, argue that it may be virtually impossible to identify truly isolated galaxies without deep multi-spectral surveys. Since Brosch et al. (2006) found much fainter SF objects surrounding dwarf, intrinsically faint, SF isolated galaxies but not interacting with them, it is possible that SF occurs synchronously in some regions due to some external agent. One possibility could be the formation of tidal dwarf galaxies following  a major interaction, but this is not the case here since no traces of interaction were detected.

\section{Star formation along a dark matter filament?}
Zitrin \& Brosch (2008) studied a number of galaxies identified as HI sources by the preliminary ALFALFA survey observations. ALFALFA is a blind HI survey conducted with the ALFA seven-feed at Arecibo (see Giovanelli et al. 2005a, b for details). The advantage of using ALFALFA is that each detection yields not only the HI content of a source but also its redshift (irrespective of its optical surface brightness), allowing its placement in 3D space.

\begin{figure}[tbh]
\centering{
  \includegraphics[angle=-90,width=6.3cm]{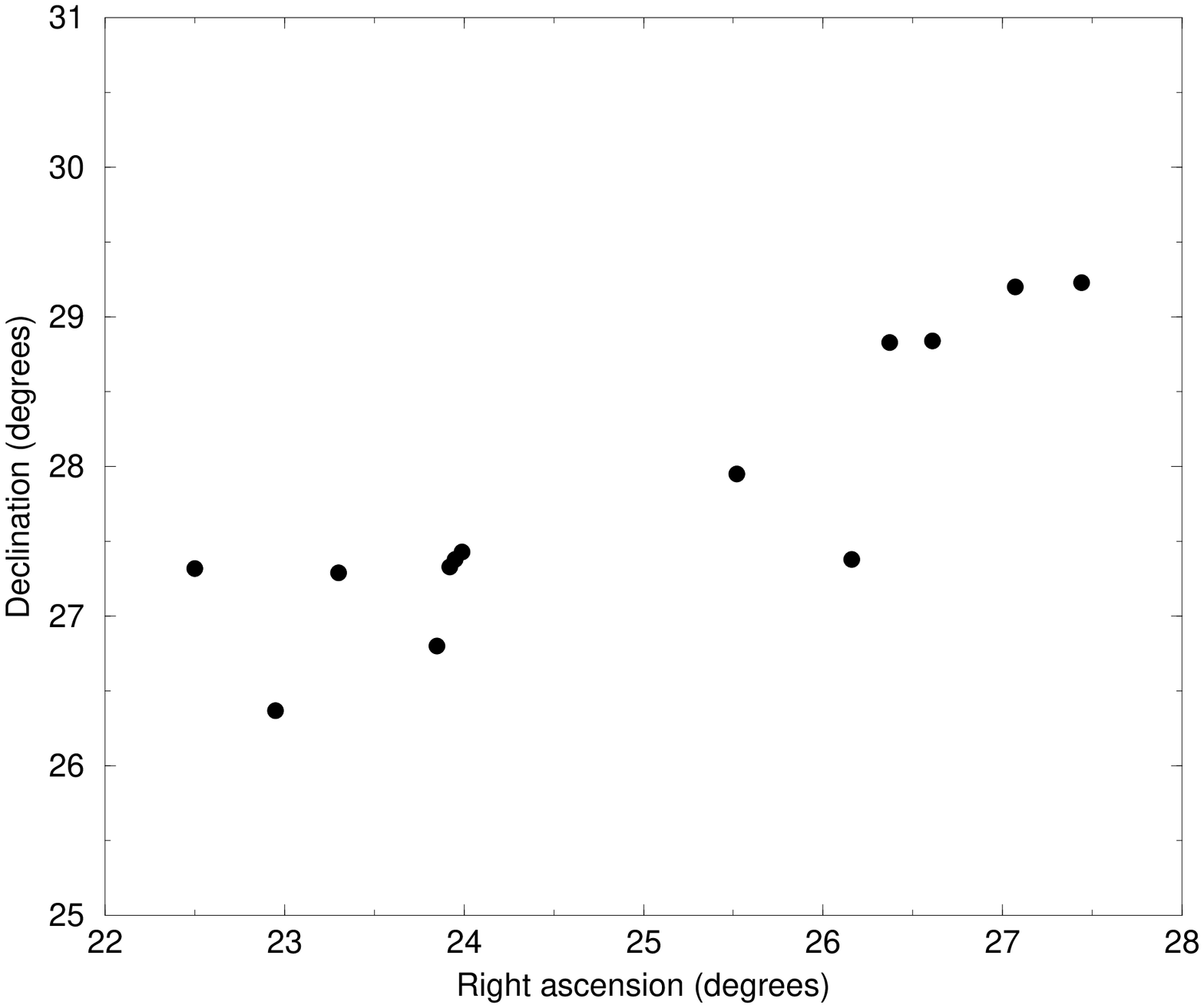}
  \includegraphics[angle=-90,width=6.3cm]{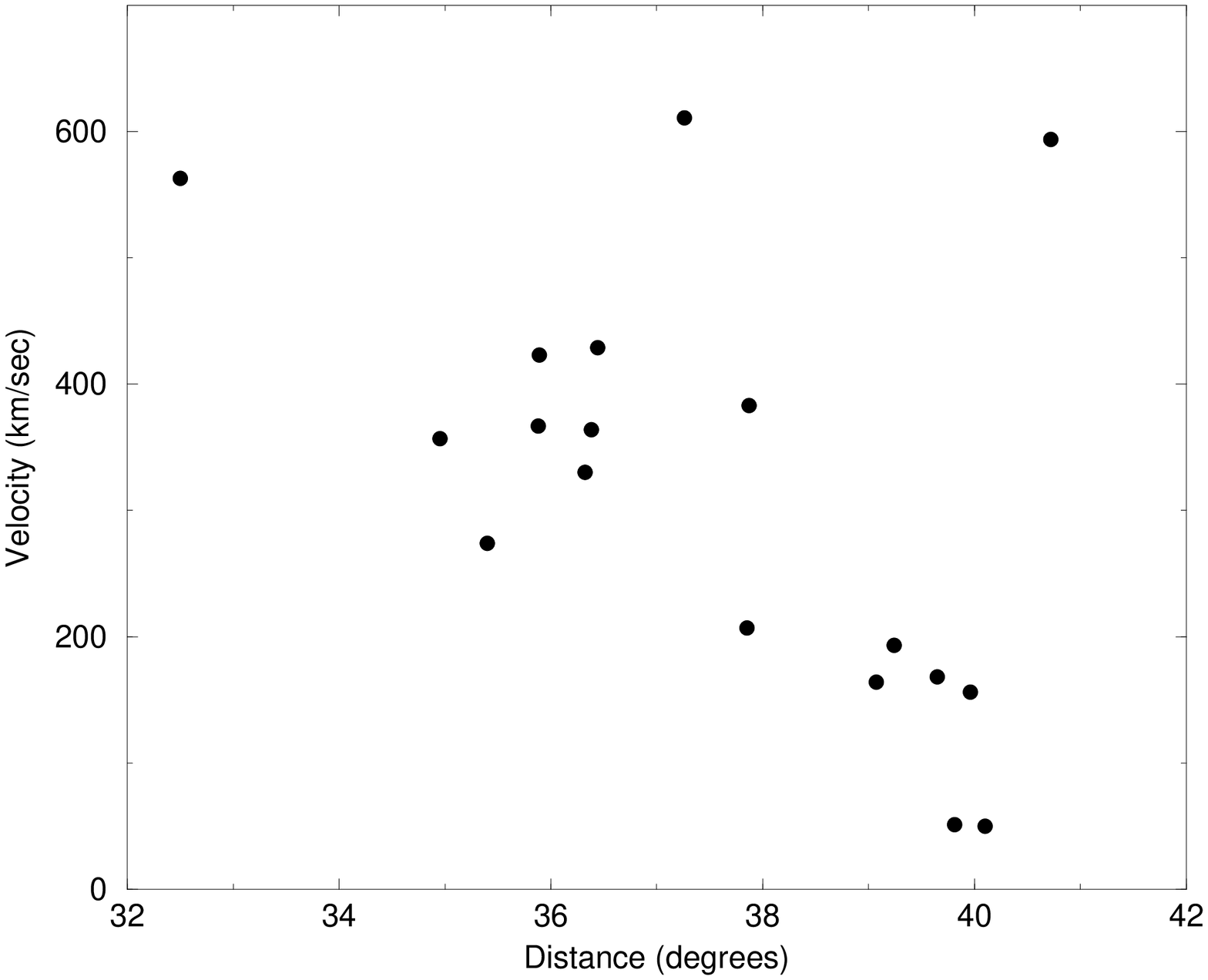}
  \caption{{\it Left panel}: Sky distribution of objects in the filament. {\it Right panel}: Position-velocity diagram for all the galaxies as projected on the sky.}
  \label{fig:gals_X_vs_Vel}}
\end{figure}

The precursor observations showed a suspicious alignment of objects at cz$\leq$600 km sec$^{-1}$ (left panel of Figure~\ref{fig:gals_X_vs_Vel}). The general region where the objects were identified is the anti-Virgo direction, in the ``Local Void'' (Tully et al. 2008). Some of the galaxies were previously associated with the classical NGC 672 and 784 galaxy groups; others had their redshifts measured for the first time by ALFALFA, and yet others were only HI clouds with no optical counterparts. The observations of Zitrin \& Brosch (2008) showed that the galaxies can be classified as ``dwarf'' (Im or Sm) with H$\alpha$ emission, indicating recent star formation. The galaxies are aligned on a $\sim$six degrees long $\sim$line and show kinematic coherence (right panel of Figure~\ref{fig:gals_X_vs_Vel}); objects at lower declinations and right ascensions have lower velocities. For this reason we label this feature as a ``galaxy filament'', a one-dimensional structure with a preferred direction in space. It appears that there might also be  a preferential alignment of the galaxies' major axes either $\sim$parallel or $\sim$perpendicular to the filament direction.

It is hard to understand why about a dozen dwarf galaxies, very distant from other objects, would be so aligned in space and would show synchronous star formation. One possibility that could explain both findings is a process of galaxy formation along a DM filament threading a nearby void. Dark matter filaments were identified in voids in N-body simulations (e.g., Gottl{\"o}ber et al. 2003). Such void filaments contain much less massive haloes than found in denser regions. Hahn et al. (2007) found that the major axes of disky galaxies in a DM filament should be approximately perpendicular to the line along which the galaxies appear to be arranged. This is only partly confirmed here, since we observe the major axes of the galaxies at PA=34$^{\circ}\pm31^{\circ}$ ($\sim$perpendicular) or 305$^{\circ}\pm12^{\circ}$ ($\sim$parallel) while the filament direction is $\sim303^{\circ}$.

Dekel \& Birnboim (2006) showed that low-mass galaxies can be built by cold gas streams accreting onto haloes below $\sim 10^{12}$ M$_{\odot}$ where the accreting intergalactic gas is not heated by a virial shock. Cold accretion can induce efficient SF in low-mass galaxies. I propose that the observational evidence argues in favor of interpreting the galaxies as located on a DM filament that is itself located in a low-galaxy-density region, and accretes intergalactic cold gas.

\section{Tentative conclusions}

I argue that finding truly isolated galaxies is an extremely difficult, almost Don Quixotean task, since locations free of luminous galaxies may harbor low-mass DM haloes. These may accrete intergalactic gas that could cause SF within the haloes, producing galaxies observed as H$\alpha$ knots or as DGs aligned on a DM filament. This implies that the nearby universe contains many small-mass haloes where SF may just be waiting to be turned on when sufficient IGM would accumulate, forming DGs. Since this happens in voids, it also follows that voids  harbor significant quantities of low-metallicity gas. If this gas is ionized and has a low column density, no current observational means can confirm or refute this possibility.


\acknowledgements 
I am grateful to Martha Haynes, Riccardo Giovanelli, and the entire ALFALFA team for providing the community with a terrific HI data set. I acknowledge long discussions and frequent interactions on some of the topics discussed here with my present and former students.



\begin{thebibliography}{}


\bibitem[Brosch \& Isaacman(1982)]{1982A&A...113..231B} Brosch, N., \& Isaacman, R.\ 1982, \aap, 113, 231




\bibitem[Brosch \& Shaviv(1982)]{1982ApJ...253..526B} Brosch, N., \& Shaviv, G.\ 1982, \apj, 253, 526

\bibitem[Brosch(1983)]{1983PhDT.........1B} Brosch, N.\ 1983, Ph.D.~Thesis, Leiden University

\bibitem[Brosch et al.(1984)]{1984A&A...135..330B} Brosch, N., Greenberg, J.~M., Rahe, J., \& Shaviv, G.\ 1984, \aap, 135, 330


\bibitem[Brosch \& Krumm(1984)]{1984A&A...132...80B} Brosch, N., \& Krumm, N.\ 1984, \aap, 132, 80



\bibitem[Brosch et al.(2006)]{2006MNRAS.368..864B} Brosch, N., Bar-Or, C.,
\& Malka, D.\ 2006, \mnras, 368, 864

\bibitem[Brosch et al.(2007)]{2007MNRAS.382.1809B} Brosch, N., et al.\
2007, \mnras, 382, 1809

\bibitem[Dekel
\& Birnboim(2006)]{2006MNRAS.368....2D} Dekel, A., \& Birnboim, Y.\ 2006, \mnras, 368, 2

\bibitem[di Matteo et
al.(2008)]{2008A&A...492...31D} di Matteo, P., Bournaud, F., Martig, M., Combes, F., Melchior, A.-L., \& Semelin, B.\ 2008, \aap, 492, 31

\bibitem[Giovanelli et al.(2005)]{2005AJ....130.2598G} Giovanelli, R., et
al.\ 2005a, \aj, 130, 2598

\bibitem[Giovanelli et al.(2005)]{2005AJ....130.2613G} Giovanelli, R., et
al.\ 2005b, \aj, 130, 2613

\bibitem[Gottl{\"o}ber et al.(2003)]{2003MNRAS.344..715G} Gottl{\"o}ber,
S., {\L}okas, E.~L., Klypin, A., \& Hoffman, Y.\ 2003, \mnras, 344, 715

\bibitem[Hahn et al.(2007)]{2007MNRAS.381...41H} Hahn, O., Carollo, C.~M.,
Porciani, C., \& Dekel, A.\ 2007, \mnras, 381, 41

\bibitem[Huchra
\& Thuan(1977)]{1977ApJ...216..694H} Huchra, J., \& Thuan, T.~X.\ 1977, \apj, 216, 694

\bibitem[Karachentseva(1973)]{1973SoSAO...8....3K} Karachentseva, V.~E.\
1973, Soobshcheniya SAO, 8, 3

\bibitem[Kellar et al.(2008)]{2008AAS...212.1906K} Kellar, J., Salzer, J.,
Wegner, G., \& Sugden, A.\ 2008, BAAS, 40, 216

\bibitem[Pisano et al.(2002)]{2002ApJS..142..161P} Pisano, D.~J., Wilcots,
E.~M., \& Liu, C.~T.\ 2002, \apjs, 142, 161

\bibitem[Sulentic et
al.(2006)]{2006A&A...449..937S} Sulentic, J.~W., et al.\ 2006, \aap, 449, 937

\bibitem[Tully et al.(2008)]{2008ApJ...676..184T} Tully, R.~B., Shaya,
E.~J., Karachentsev, I.~D., Courtois, H.~M., Kocevski, D.~D., Rizzi, L.,
\& Peel, A.\ 2008, \apj, 676, 184

\bibitem[Zitrin \& Brosch(2008)]{2008MNRAS.390..408Z} Zitrin, A., \& Brosch, N.\ 2008, \mnras, 390, 408


\end{thebibliography}
\end{document}